\documentclass[12pt, a4paper]{article}

\usepackage{amssymb, amsmath}  
\usepackage{fullpage}
\usepackage{amsfonts}
\usepackage{graphicx}
\usepackage{mathrsfs}
\usepackage{latexsym}	
\usepackage{oldgerm}
\usepackage{color}
\usepackage{cite}
\usepackage{float}
\usepackage{epstopdf}
\usepackage{cite} 

\newtheorem{definition}{Definition}[section]

\newtheorem{theorem}{Theorem}[section]
\newtheorem{example}{Example}[section]

\newcommand{\bdf}{\begin{definition}}
\newcommand{\edf}{\end{definition}}
\newcommand{\bthm}{\begin{theorem}}
\newcommand{\ethm}{\end{theorem}}
\newcommand{\bex}{\begin{example}}
\newcommand{\eex}{\end{example}}

\newtheorem{prop}{PROPOSITION}[section]
\newcommand{\bprop}{\begin{prop}}
\newcommand{\eprop}{\end{prop}}
\newcommand{\bdis}{\begin{displaymath}}
\newcommand{\edis}{\end{displaymath}}
\newcommand{\beqn}{\begin{equation}}
\newcommand{\eeqn}{\end{equation}}

\newcommand{\bgb}{\begin{gbox}}
\newcommand{\egb}{\end{gbox}}
\newcommand{\bdb}{\begin{darkbox}}
\newcommand{\edb}{\end{darkbox}}

\newcommand{\bsplit}{\begin{split}}
\newcommand{\esplit}{\end{split}}

\usepackage{xcolor}
\usepackage{amsthm}
\usepackage{framed}
\usepackage{authblk}

\colorlet{shadecolor}{gray!15}
\newenvironment{gbox}
  {\begin{shaded}}
  {\end{shaded}}

\title{The Stability of the Higgs vacuum inside a Schwarzschild Black Hole.} 

\date{\today}

\author{ Paul G. Abel \\
{\small  School of Physics \& Astronomy,} \\
{\small  University of Leicester, Leicester UK.} \\
{\small  LE1 7RH.} \\
{\small {Email:} pga3@le.ac.uk}}

\begin{document}

\maketitle

\begin{abstract}
I investigate the behaviour of the Higgs field inside a Schwarzschild black hole, including a Planck-suppressed coupling between the Higgs field and the Kretschmann scalar, the leading curvature invariant in a Ricci-flat geometry. Because the radial coordinate is timelike in the interior, the Higgs condensate evolves in time rather than tracking the minimum of its potential. Near the singularity the interior field equation reduces to Bessel's equation, with solution $H=vJ_0(x)$, where $x\propto\sqrt\xi\,t_p/\tau$ and $\tau$ is the proper time to the singularity. Contrary to static analyses, the field remains frozen in its broken phase while the potential changes shape, departing from its vacuum value only when the curvature becomes Planckian, independently of the black-hole mass. Including the one-loop correction from Higgs vacuum fluctuations yields the same Bessel equation with a source term, which I solve exactly; the condensate is reduced by at most about ten per cent. The electroweak symmetry therefore remains broken throughout the interior within the regime of validity of effective field theory.

\end{abstract}

\section{Introduction}
The discovery of the Higgs boson completed the Standard Model of particle physics and confirmed the mechanism responsible for electroweak symmetry breaking. Through spontaneous symmetry breaking, the Higgs field acquires a non-zero vacuum expectation value, $v\simeq246\,\mathrm{GeV}$ \cite{higgsvacexpvalue}, which generates the masses of the weak gauge bosons and the fermions while leaving the photon massless. Although the Higgs mechanism has been studied extensively in flat spacetime and at finite temperature, considerably less is known about its behaviour in strong gravitational fields, where spacetime curvature may itself modify the structure of the Higgs vacuum. Several results already indicate that gravity can play a significant role: black holes can act as nucleation sites that catalyse the decay of the metastable electroweak vacuum \cite{PhysRevLett.115.071303}, vacuum fluctuations of the Higgs field can have important gravitational consequences \cite{PhysRevD.98.123509}, and vacuum polarisation in the vicinity of a black hole can restore a spontaneously broken symmetry close to the horizon \cite{QFL}.
\\

Quantum field theory in curved spacetime provides the natural framework for studying quantum fields on a classical gravitational background \cite{Birrell_Davies_1982, qftcurved}. In this approach the geometry modifies the dynamics of matter fields, while the gravitational field itself is treated classically. Since Hawking's demonstration that black holes radiate \cite{Hawking}, it has been clear that strong curvature can have dramatic quantum consequences, and the behaviour of quantum fields near and inside black holes remains an active area of research.
\\

Within this framework, curvature corrections to the dynamics of scalar fields are expected to arise naturally, either through non-minimal couplings or through higher-dimensional operators in an effective field theory valid below some ultraviolet scale. Understanding how such corrections affect the Higgs vacuum is relevant both to black-hole physics and, more generally, to the behaviour of matter in the strong-field regime of gravity.
\\

Most previous studies have considered the non-minimal interaction $\xi RH^2$, where $R$ is the Ricci scalar; see, for example, \cite{GIALAMAS2023137885}. Such couplings play a central role in Higgs inflation \cite{Bezrukov2008} and in scalar-field dynamics in cosmological spacetimes. The Schwarzschild solution, however, is Ricci-flat, $R_{\mu\nu}=0$, so all couplings to the Ricci scalar or Ricci tensor vanish identically. They cannot modify the Higgs vacuum either outside or inside a Schwarzschild black hole.
\\

Any gravitational modification of the Higgs sector in a Ricci-flat geometry must therefore arise from the Riemann tensor. In such a spacetime the Riemann tensor reduces to the Weyl tensor, and the only non-vanishing quadratic scalar invariant is the Kretschmann scalar,
\beqn
K=R_{\mu\nu\rho\sigma}R^{\mu\nu\rho\sigma}=\frac{12r_s^2}{r^6},
\eeqn
which provides a coordinate-independent measure of the curvature and grows without bound towards the singularity. It is therefore the natural quantity through which gravity can couple to the Higgs field in a Ricci-flat background. Motivated by effective field theory, we introduce the interaction
\beqn
\mathcal L_{\rm grav}=-\xi\,\ell_p^2\,K\,H^2 ,
\eeqn
where $\xi$ is a dimensionless coupling and $\ell_p$ is the Planck length. Since $K$ has mass dimension four, the factor $\ell_p^2$ supplies the suppression expected for the leading curvature correction generated by quantum gravity, and this is the lowest-dimension curvature--Higgs operator that is non-vanishing in a Ricci-flat spacetime. Couplings of matter fields to the Kretschmann scalar have also been considered elsewhere, for example in the study of Higgs-induced spectroscopic shifts in strong gravitational fields \cite{PhysRevD.82.065008}.
\\

The interaction contributes a curvature-dependent term $2\xi\ell_p^2K$ to the effective Higgs mass squared. Outside the black hole this term is negligible, but it grows rapidly in the interior, where it eventually exceeds the negative mass term responsible for symmetry breaking. A static analysis, in which the Higgs field is assumed to follow the minimum of its effective potential, then predicts curvature-induced restoration of electroweak symmetry at a critical curvature $K_c=2\lambda v^2/(\xi\ell_p^2)$ \cite{Abel2026}. In this sense the curvature plays a role analogous to temperature in thermal symmetry restoration.
\\

Inside the event horizon, however, the radial coordinate is timelike, and the interior is a time-dependent, homogeneous cosmology evolving towards the singularity. The Higgs field must therefore be evolved dynamically, and there is no a priori reason for it to track the minimum of a potential that is itself changing in time. In this paper I investigate the dynamics of the Higgs field inside a Schwarzschild black hole in the presence of the Kretschmann coupling. I derive the field equation for the Higgs condensate in the interior and show that, in the region where the curvature becomes significant, it reduces to Bessel's equation, whose solution can be obtained analytically. I then compute the one-loop correction to this solution arising from vacuum fluctuations of the Higgs field, treating it perturbatively about the classical solution.
\\

The results show that the Higgs vacuum is remarkably robust. The condensate is described by $H=vJ_0(x)$, with $x\propto\sqrt\xi\,t_p/\tau$, where $\tau$ is the proper time remaining before the singularity. The field cannot respond to the change in its potential: it remains frozen in its broken phase until the curvature becomes Planckian, at a proper time of order $\sqrt\xi\,t_p$ before the singularity, independently of the mass of the black hole. The one-loop correction from vacuum fluctuations reduces the condensate by at most about ten per cent and does not alter this conclusion. Within the regime of validity of effective field theory the electroweak symmetry therefore remains broken throughout the black-hole interior, and any restoration of the symmetry must arise in the Planckian regime, where quantum-gravitational effects are expected to dominate.
\\

The results suggest the the Higgs field vacuum state is actually rather robust.  It remains in its broken phase until a distance of order a Planck length from the singularity.  At this point one would expect quantum gravitational effects to dominate, so if there is to be any restoration of the Higgs vacuum state, it must be due to quantum gravitational effects since we effectively show the Higgs field remains in the broken phase up until this point.

\section{The geometry of the black hole interior}

It will be useful to discuss the interior geometry of the Schwarzschild black hole as we shall need to modify it in a way that makes it more appropriate to our problem.  The Schwarzschild metric can be written in the form:
\beqn
ds^2=-\left(1-\frac{r_s}{r} \right)dt^2+\left(1-\frac{r_s}{r} \right)^{-1}dr^2+r^2d\Omega^2
\eeqn
where the function
\beqn
f=1-\frac{r_s}{r},
\eeqn
and $r_s=2GM$ is the usual Schwarzschild event horizon radius.  The primary issue now is that $r<r_s$ and so the factor
\bdis
1-\frac{r_s}{r}<0.
\edis
which means that the $t$ and $r$ coordinates have effectively swapped roles.  We introduce a function
\beqn
f=\frac{r_s}{r}-1
\eeqn
and we see that, in the interior, $f>0$.  The metric is now
\beqn
ds^2= -f^{-1}dr^2+f dt^2+r^2d\Omega^2.
\eeqn
The coefficient of $dr^2$ is now negative which makes $r$ timelike and $dt^2$ is now the spacelike coordinate.  We can make a further change which will the metric much more convenient for our purposes.  We shall introduce the spatial coordinate
\beqn
\sigma=\ln\left(\frac{r}{r_s-r} \right)
\label{sigma}
\eeqn

We see that at the horizon, $\sigma\rightarrow \infty$, and at the singularity $\sigma\rightarrow -\infty$, and now the metric for the interior  becomes
\beqn
ds^2=-N^2d\sigma^2+e^{-\sigma}dz^2+r^2d\Omega^2
\label{metric}
\eeqn
where
\beqn
N^2=\frac{r^3(r_s-r)}{r_s^2}=\frac{r_s^2e^{3\sigma}}{(1+e^\sigma)^4}
\label{N}
\eeqn
and we have let $t\rightarrow z$ which is now our spacelike coordinate.   The metric written in this form effectively means that for an observer who has fixed coordinates $(z,\theta,\phi)$ that
\bdis
d\tau^2=-N^2d\sigma^2
\edis
so in this respect, $N$ can be thought of as a conversion factor; converting $d\sigma$ intervals into $d\tau$ ones.  
\\

There are additional advantages of writing the metric in this form.  Firstly, the topology of the interior is now very convenient; the interior is stretched over $\mathbb R$ since $0<r<r_s$ is now mapped to $-\infty<\sigma<\infty$.  It also has the advantage of giving the spatial components of the metric a simple form whereby $\sigma$ is a harmonic time coordinate since $\Box\sigma=0$.
\\

In the coordinates $(\sigma,z,\theta,\phi)$, the black-hole interior has the topology $\mathbb{R}\times\mathbb{R}\times \mathbb S^2$. The first factor is the time coordinate $\sigma\in(-\infty,\infty)$. Each hypersurface of constant $\sigma$ is a spatial slice with topology $\mathbb{R}\times \mathbb S^2$: an infinite line parametrised by $z\in(-\infty,\infty)$, times a round two-sphere of radius $r(\sigma)=r_se^{\sigma}/(1+e^{\sigma})$. These slices are homogeneous but anisotropic, so the interior is a Kantowski–Sachs cosmology. Its evolution runs from the event horizon, reached as $\sigma\to+\infty$, to the curvature singularity, reached as $\sigma\to-\infty$. Neither the horizon nor the singularity belongs to the coordinate patch. The horizon is a regular null boundary that can be crossed into the exterior region. The singularity is a spacelike boundary in the future of every observer. As $\sigma$ decreases, the two-sphere contracts to zero radius ($r\to0$), while the $z$-direction stretches without bound ($g_{zz}=e^{-\sigma}\to\infty$). The singularity is therefore a spacelike “cigar” crunch, in which each spatial slice collapses to an infinitely long line.

\section{The interior Higss-Schwazschild field equation}

We now derive the Higgs-Schwarzschild field equation.  We start with the Lagrangian for the system:
\beqn
\mathcal L=\sqrt{-g}\left[ -\frac{1}{2}g^{\mu\nu}\partial_\mu H\partial_\nu H-V(H)\right]
\eeqn 
where the potential term is, as explained earlier now the standard Higgs potential with the coupling to gravity represented by the Kretchmann scalar
\beqn
V(H)=\lambda(H^2-v^2)^2+\xi\ell_p^2KH^2.
\label{potential}
\eeqn
Using the Euler-Lagrange equations we obtain the Higgs Schwarzschild interior field equation:
\beqn
\frac{\partial^2H}{\partial\sigma^2}-N^2e^\sigma\frac{\partial^2H}{\partial z^2}-\frac{N^2}{r^2}\Delta_{\mathbb S^{2}}H+N^2[4\lambda H(H^2-v^2)+2\xi\ell_p^2KH]=0.
\label{HS1}
\eeqn

Here we have used the notation $\Delta_{\mathbb S^2}$ to represent the Laplacian on the unit two-sphere:
\beqn
\Delta_{\mathbb S^2}f=\frac{1}{\sin(\theta)}\frac{\partial}{\partial \theta}\left(\sin(\theta )\frac{\partial f}{\partial \theta}\right)+\frac{1}{\sin^2(\theta)}\frac{\partial^2 f}{\partial \phi^2}.
\eeqn
\\
We can now simplify the field equation. As noted above, inside the black hole $r$ and $t$ exchange roles: $r$ (equivalently $\sigma$) is the time coordinate, and $t$, relabelled $z$, is spatial. On a slice of constant $\sigma$, the metric components are independent of $z$, and the angular part is the round metric of a two-sphere of radius $r(\sigma)$. Each slice is therefore homogeneous: translations along $z$ and rotations of the sphere leave the geometry unchanged. A field sharing these symmetries depends on time alone, and we take the Higgs condensate to be $H=H(\sigma)$. This is a consistent truncation of the full field equation, since all its coefficients depend only on $\sigma$. Inhomogeneous fluctuations about this background are treated separately.  Since $H=H(\sigma)$ then (\ref{HS1}) reduced to:
\beqn
\frac{d^2H}{d\sigma^2}+N^2\left[4\lambda H(H^2-v^2)+\frac{24\xi\ell_p^2r_s^2H}{r^6}\right]=0.
\eeqn
\label{HS1}
The extrema of the potential follow from $V'(H)=0$, which gives $H=0$ together with $H^2=v^2-\xi\ell_p^2K/(2\lambda)$. The non-trivial minima exist only for $K<K_c$, where
\beqn
K_c=\frac{2\lambda v^2}{\xi\ell_p^2}
\eeqn
Equivalently, $K_c$ is defined by $V''(0)=-4\lambda v^2+2\xi\ell_p^2K=0$: at this value the curvature contribution exactly cancels the negative curvature of the Higgs potential at the origin. For $K>K_c$ the potential has a single minimum at $H=0$. Inside the black hole, $K$ increases monotonically with time, so $K_c$ defines a critical time $\sigma_c$ at which the potential changes from the broken to the symmetric shape. Whether the field actually follows this change is a separate, dynamical question which we will go on to address.
\\

The associated critical radius occurs when $K=K_C$:
\beqn
\frac{r_c}{rs}=\left( \frac{6\xi\ell_p^2}{\lambda v^2 r_s^4}\right)^{1/6}
\eeqn
and using appropriate values it can be shown that $r_c\sim3\times 10^{20}\xi^{1/6}r_s$, so $r_c\ll r_s$ for any astrophysical black hole.  In the interior, $r<r_s$ and so 
\beqn
e^\sigma \sim \frac{r}{r_s}\ll 1,
\eeqn
and as a result we may say that
\beqn
N^2\sim r_s^2e^{3\sigma},
\eeqn
Similarly we may approximate the Kretschmann scalar:
\beqn
K\sim\frac{12}{r_s^4}e^{-6\sigma}
\eeqn
and so we can write the interior Higgs-Schwarzschild field equation as:
\beqn
\frac{d^2H}{d\sigma^2}+ae^{3\sigma}H\left(\frac{H^2}{v^2}-1\right)+be^{-3\sigma}H=0,
\eeqn
where the coefficients
\beqn
a=4\lambda v^2r_s^2 \ \ \mathrm{and} \ \ b=\frac{24\xi\ell_p^2}{r_s^2}.
\eeqn
As the black hole interior evolves from the event horizon toward the singularity, the curvature $K$ grows, and at a critical time $\sigma_c$ the shape of the potential changes. Whether $H=0$ is a maximum or a minimum is determined by the sign of $V''(0;\sigma)$. Before $\sigma_c$, $V''(0;\sigma)<0$: $H=0$ is a maximum, and the minima lie at $H=\pm v_{\rm eff}\neq0$, so the potential has the broken-symmetry form. After $\sigma_c$, $V''(0;\sigma)>0$, and $H=0$ is the only minimum, so the potential has the symmetric form. The critical time is therefore defined by $V''(0;\sigma_c)=0$.  We can determine $\sigma_c$ from:
\bdis
V''(H;\sigma)=\frac{3ae^{3\sigma} H^3}{v^2}-ae^{3\sigma}H+be^{-3\sigma}H,
\edis
by solving $V''(0;\sigma)=0$ the critical time is:
\beqn
\sigma_c=\frac 1 6 \ln\left(\frac{b}{a}\right).
\eeqn
It is convenient to measure time from this critical moment, and so we define
\beqn
s=\sigma-\sigma_c ,
\eeqn
so that $s=0$ corresponds to $\sigma=\sigma_c$. Since the interior evolves toward decreasing $\sigma$, the region $s>0$ lies before the critical time, closer to the event horizon, while $s<0$ lies after it, closer to the singularity. As we shall see, this shift places the Higgs and curvature terms on an equal footing and reduces the equation of motion to a form depending on a single parameter:
\beqn
\frac{d^2H}{ds^2}+\Omega^2\left[e^{3s}\left(\frac{H^2}{v^2}-1\right)H+e^{-3s}H \right]=0
\label{HS3}
\eeqn
and where we have defined the quantity
\beqn
\Omega^2=\sqrt{ab}=\sqrt{12\xi M^2_H\ell_p^2}.
\eeqn
where we have used the fact that $m_H^2=8\lambda v^2$.  Again substituting reasonable values into this expressions shows that $\Omega$ is a small quantity.

\section{Solution of the Higgs-Schwarzschild field equation in the interior}

The field equation (\ref{HS3}) is non-linear with time-dependent coefficients, so no closed-form solution is available. It can, however, be solved approximately by examining when each of the two terms driving the field, the self-interaction and the curvature coupling, is important.  For $r<r_s$:
\bdis
\frac{d^2H}{ds^2}=-\Omega^2e^{3s}\left(\frac{H^2}{v^2}-1\right)H-\Omega^2e^{-3s}H .
\edis
Each coefficient changes by a factor $e^3\approx20$ per unit of $s$, so each term acts on the field for an interval $\Delta s\sim1$, during which it changes $H$ by an amount of order (coefficient)$\times H$. A term therefore has an appreciable effect on the field only when its coefficient is of order unity or larger. For the self-interaction term this threshold is reached at $s=s_*$, where $\Omega^2e^{3s_*}=1$, giving
\beqn
s_*=\frac{2}{3}\ln\left(\frac{1}{\Omega}\right) .
\eeqn
For the curvature term the threshold $\Omega^2e^{-3s}=1$ is reached at $s=-s_*$. Since $\Omega\ll1$, $s_*>0$, and the interior divides into three regions:
\begin{itemize}
	\item \emph{Region I} ($s>s_*$): extending from the event horizon to $s_*$, before the critical time. The self-interaction term dominates and the curvature term is negligible.
	\item \emph{Region II} ($-s_*<s<s_*$): neither term has an appreciable effect on the field. The critical time $\sigma_c$ ($s=0$) lies at the centre of this region.
	\item \emph{Region III} ($s<-s_*$): after the critical time, close to the singularity. The curvature term dominates and the self-interaction term is negligible.
\end{itemize}
In region I, the interaction term dominates and the field equation is
\beqn
\frac{d^2H}{ds^2}=-\omega^2e^{3s}\left(\frac{H^2}{v^2}-1\right)H
\eeqn
and in this case the solution is $H=v$.  
\\

In Regions II and III ($s<s_*$), the self-interaction term is negligible. Its coefficient satisfies $\Omega^2e^{3s}\ll1$, and since $|H^2/v^2-1|\le1$ for $|H|\le v$, the term is bounded by $\Omega^2e^{3s}|H|\ll|H|$. Dropping it, the field equation becomes
\beqn
\frac{d^2H}{ds^2}+\Omega^2e^{-3s}H=0 .
\label{H-C}
\eeqn
We make the change of variable
\beqn
x=\frac{2\Omega}{3}e^{-3s/2},
\eeqn
so that $dx/ds=-\tfrac32x$ and $\Omega^2e^{-3s}=\tfrac94x^2$. Substituting into (\ref{H-C}) gives
\beqn
x^2\frac{d^2H}{dx^2}+x\frac{dH}{dx}+x^2H=0 ,
\eeqn
which is Bessel’s equation of order zero, with general solution \cite{Gradshteyn}
\beqn
H(x)=C_1J_0(x)+C_2Y_0(x) ,
\label{BH}
\eeqn
where $C_1$ and $C_2$ are constants. $J_0$ is the Bessel function of the first kind,
\beqn
J_0(x)=\sum_{k=0}^\infty\frac{(-1)^k}{(k!)^2}\left(\frac x2\right)^{2k} ,
\eeqn
and $Y_0$ is the Bessel function of the second kind,
\beqn
Y_0(x)=\frac{2}{\pi}\left[\left(\ln\frac x2+\gamma\right)J_0(x)-\sum_{k=1}^\infty\frac{(-1)^k\mathcal H_k}{(k!)^2}\left(\frac x2\right)^{2k}\right] ,
\eeqn
with $\gamma\approx0.5772$ the Euler–Mascheroni constant and $\mathcal H_k=\sum_{j=1}^k 1/j$ the $k$-th harmonic number. As time runs toward the singularity, $x$ increases from $x\to0$ at the horizon to $x\to\infty$ at the singularity.
\\

The constants $C_1$ and $C_2$ are fixed by matching to Region I. The Region I equation, in which the curvature term is dropped, is valid for $s>-s_*$, and its solution is $H=v$, $H'=0$. The Bessel equation (\ref{H-C}) is valid for $s<s_*$. Both approximations therefore hold in the overlap $-s_*<s<s_*$, where they must describe the same function. We choose a matching point $s_0$ in this interval, with $x_0=x(s_0)$, and require
\beqn
H(s_0)=v, \qquad \left.\frac{dH}{ds}\right\vert_{s_0}=0 .
\eeqn
Using the chain rule and the identities $J_0'(x)=-J_1(x)$ and $Y_0'(x)=-Y_1(x)$,
\bdis
\frac{dH}{ds}=\frac{dx}{ds}\frac{dH}{dx}=\frac32x\left[C_1J_1(x)+C_2Y_1(x)\right] ,
\edis
so the matching conditions become
\begin{eqnarray}
	C_1J_0(x_0)+C_2Y_0(x_0) & = & v , \\
	C_1J_1(x_0)+C_2Y_1(x_0) & = & 0 .
\end{eqnarray}
Solving these, and using the Wronskian $J_1(x)Y_0(x)-J_0(x)Y_1(x)=2/(\pi x)$, gives
\beqn
C_1=-\frac{\pi x_0}{2}Y_1(x_0),v , \qquad C_2=\frac{\pi x_0}{2}J_1(x_0),v .
\label{C1C2}
\eeqn
\\

We take the matching point to be $s_0=0$, the midpoint of the interval $(-s_*,s_*)$. At this point both of the neglected terms have the same size, $\Omega^2$, so the two approximations are equally accurate there. Then
\beqn
x_0=\frac{2\Omega}{3}\ll1 ,
\eeqn
and we may use the small-argument forms
\beqn
J_1(x)=\frac{x}{2}-\frac{x^3}{16}+\dots\simeq\frac{x}{2}, \qquad Y_1(x)=-Y_0'(x)\simeq-\frac{2}{\pi x} .
\eeqn
Substituting these into (\ref{C1C2}) gives
\beqn
C_1\simeq v , \qquad C_2\simeq\frac{\pi}{9}\Omega^2v\approx1.2\times10^{-17}\sqrt{\xi},v .
\eeqn
The $C_2$ term is therefore a correction of relative order $\Omega^2$, the same order as the terms already neglected in deriving the approximate equations, and we set $C_2=0$ consistently. The solution is
\beqn
H=vJ_0\left(\frac{2\Omega}{3}e^{-3s/2}\right) .
\label{HR2-3}
\eeqn
This single expression describes all three regions.

In Region I, which extends from the event horizon to $s_*$, $x<10^{-17}$ and $J_0(x)=1$ to high accuracy, so $H=v$: the field sits in its broken-symmetry minimum.

In Region II, which spans equal intervals before and after the critical time $\sigma_c$, we still have $x\ll1$, so $H\simeq v(1-x^2/4)\simeq v$. The field remains at $v$ even though, after $\sigma_c$, the minimum of the potential has moved to $H=0$.

In Region III, close to the singularity, the field begins to move once $x\sim1$, first crossing zero at $x\simeq2.405$. For $x\gg1$, the large-argument form of $J_0$ gives
\beqn
H\simeq v\sqrt{\frac{2}{\pi x}}\cos\left(x-\frac{\pi}{4}\right) ,
\eeqn
so the field oscillates with an amplitude that decays as $x^{-1/2}\propto\tau^{1/2}$ as the singularity is approached.

Since $x\simeq2.18\sqrt\xi\,t_p/\tau$, we conclude that from the event horizon, through the critical time, and down to a proper time $\tau\sim\sqrt\xi\,t_p$ before the singularity (where $K\sim0.1\,\xi^{-2}\ell_p^{-4}$, Planckian for $\xi\sim1$), the Higgs field remains in its broken phase, $H=v$. Only then does it ring down towards zero. At these curvatures the effective description is expected to fail, and quantum-gravitational effects may well alter this behaviour. Within the classical theory, however, the Higgs field remains in its broken phase throughout essentially the entire interior. Quantum corrections from the Standard Model fields themselves are not Planck-suppressed, and may modify this conclusion well before the Planck regime. We turn to these in the next section.

\section{Quantum Fluctuations at the one-loop level}

The classical solution (\ref{HR2-3}) shows that the Higgs condensate remains in its broken phase, $H\simeq v$, throughout Region~II. We now determine how this conclusion is modified by vacuum fluctuations of the Higgs field at one loop.
\\

\subsection{Higss Field in the One Loop}
The field equation 
\bdis
\hat H''+N^2V'(\hat H)=0 
\label{fe1}
\edis
holds for the field operator. We decompose the field operator:
\beqn
\hat H=H_0+H_1+\delta\hat H ,
\eeqn
where $H_0=vJ_0(x)$ is the classical solution found in the previous section and satisfies (\ref{fe1}). $H_1$ is the correction to the condensate induced by quantum fluctuations, and $\delta\hat H$ is the fluctuation, with $\langle\delta\hat H\rangle=0$. Since $V'$ is cubic in $H$, its expansion about $H_0$ is,
\beqn
V'(H_0+H_1+\delta H)=V'(H_0)+V''(H_0)(H_1+\delta H)+\tfrac12V'''(H_0)(H_1+\delta H)^2+\tfrac16V''''(H_0)(H_1+\delta H)^3 .
\eeqn
Taking the expectation value, terms odd in $\delta H$ vanish, since the fluctuations are Gaussian with zero mean. To leading order the condensate correction obeys
\beqn
H_1''+N^2V''(H_0)\,H_1=-N^2\,\mathcal K\,\langle\delta H^2\rangle\,H_0 , ,
\label{H1eq}
\eeqn
where we have let 
\beqn
\mathcal K=\frac{V'''(H_0)}{2H_0}=12\lambda
\eeqn
and $\langle\delta H^2\rangle$ denotes the renormalised variance. The neglected terms, those in $H_1^2$, $H_1^3$ and $H_1\langle\delta H^2\rangle$, are of second order in the fluctuations and the curvature coupling is quadratic in $H$ and does not contribute to $V'''$.
\\

Equation (\ref{H1eq}) includes only the radial fluctuation. Writing the Higgs doublet in terms of four real fields, with 
\beqn
H^2=2\Phi^\dagger\Phi=\phi_1^2+\phi_2^2+\phi_3^2+\phi_4^2, 
\eeqn
the potential term becomes
\beqn 
V=\lambda(\sum_a\phi_a^2-v^2)^2+\xi\ell_p^2K\sum_a\phi_a^2.
\eeqn 
Taking the condensate along $\phi_1$ and averaging $\partial V/\partial\phi_1$ over the radial fluctuation and the three Goldstone modes $\pi_a$ gives the source
\beqn
12\lambda\,\langle\delta H^2\rangle\,H_0+4\lambda\sum_{a=1}^{3}\langle\pi_a^2\rangle\,H_0 .
\eeqn
In Region~II the radial and Goldstone fluctuations are effectively massless and have identical gradient terms, so their variances coincide and $\mathcal K=24\lambda$. The Goldstone contribution is gauge dependent, and a gauge-invariant result requires the inclusion of the gauge bosons; the radial contribution, $\mathcal K=12\lambda$, is gauge independent.

\subsection{The variance near the singularity}

In Region~II the fluctuations are effectively massless, $m_H\tau\ll1$, where $\tau$ is the proper time to the singularity. The renormalised variance of a massless scalar near the Schwarzschild singularity has been computed in \cite{JiangJiang}, with the result that 
\beqn
\langle\delta H^2\rangle=\frac{qM}{r^3}, \ q\simeq7.18\times10^{-4}. 
\eeqn
The leading coefficient is the same in any Hadamard state, and in particular in the Unruh state appropriate to a black hole formed by collapse. Near the singularity $\tau=\tfrac23r^{3/2}/\sqrt{r_s}$ and $N^2=\tfrac94\tau^2$, so that
\beqn
\langle\delta H^2\rangle=\frac{C}{\tau^2},\qquad C=\frac{2q}{9}\simeq1.6\times10^{-4},
\label{varC}
\eeqn
and the source in (\ref{H1eq}) takes the form $-\mu^2H_0$, with constant coefficient
\beqn
\mu^2=N^2\,\mathcal K\,\langle\delta H^2\rangle=\tfrac94\,\mathcal K\,C\simeq
\begin{cases}1.4\times10^{-4}, & \mathcal K=12\lambda,\\ 2.8\times10^{-4}, & \mathcal K=24\lambda.\end{cases}
\label{mu2}
\eeqn
In Region~I, where $m_H\tau>1$, the fluctuations are massive and their renormalised variance is suppressed, so the source is negligible and $H_1=0$ there.

\subsection{Reduction to Bessel's equation with a source}

In the variables of Section 4, the operator in (\ref{H1eq}) is
\beqn
N^2V''(H_0)=\Omega^2e^{3s}\left(\frac{3H_0^2}{v^2}-1\right)+\Omega^2e^{-3s}.
\eeqn
Since $|H_0|\le v$, the first term is bounded by $2\Omega^2e^{3s}$, which is negligible for $s<s_*$; this is the same approximation used in obtaining $H_0$. Equation (\ref{H1eq}) therefore reduces to
\beqn
H_1''+\Omega^2e^{-3s}H_1=-\mu^2v\,J_0(x),\qquad s<s_* .
\label{H1s}
\eeqn
Since $H_1$ vanishes for $s>s_*$, and $H_1$ and $H_1'$ are continuous, the initial conditions are
\beqn
H_1(s_*)=H_1'(s_*)=0 .
\label{H1ic}
\eeqn

As in section 4, we now change variable to $x=\tfrac{2\Omega}{3}e^{-3s/2}$ and (\ref{H1s}) becomes Bessel's equation of order zero with a source term,
\beqn
\frac{d^2H_1}{dx^2}+\frac1x\frac{dH_1}{dx}+H_1=F(x),
\label{BesselF}
\eeqn
and the source term is
\beqn
\qquad F(x)=-\frac{c\,J_0(x)}{x^2}, 
\eeqn
and $c=\frac{4\mu^2v}{9}=\mathcal K\,C\,v$.  The initial time $s=s_*$ corresponds to $x_*=\tfrac23\Omega^2$, using $e^{3s_*}=\Omega^{-2}$, and (\ref{H1ic}) becomes $H_1(x_*)=dH_1/dx|_{x_*}=0$.
\\

The homogeneous solutions of (\ref{BesselF}) are $J_0(x)$ and $Y_0(x)$, with Wronskian
\beqn 
W=J_0Y_0'-J_0'Y_0=2/(\pi x). 
\eeqn
The solution satisfying the initial conditions follows from the standard variation of parameters formula,
\beqn
H_1(x)=-J_0(x)\int_{x_*}^{x}\frac{Y_0F}{W}\,dx'+Y_0(x)\int_{x_*}^{x}\frac{J_0F}{W}\,dx' ,
\eeqn
which, with $F/W=-\tfrac{\pi c}{2}J_0/x$, gives
\beqn
H_1(x)=\frac{\pi c}{2}\left[J_0(x)\int_{x_*}^{x}\frac{J_0(x')\,Y_0(x')}{x'}\,dx'-Y_0(x)\int_{x_*}^{x}\frac{J_0(x')^2}{x'}\,dx'\right].
\label{H1sol}
\eeqn

Throughout Region~II, $x\ll1$. Using $J_0\simeq1$ and $Y_0\simeq\tfrac2\pi\alpha(x)$, with $\alpha(x)=\ln(x/2)+\gamma$, the two integrals evaluate to $\tfrac1\pi(\alpha^2-\alpha_*^2)$ and $\alpha-\alpha_*$ respectively, where $\alpha_*=\alpha(x_*)$. This gives
\beqn
H_1\simeq-\frac c2\,(\alpha-\alpha_*)^2=-\frac{\mu^2v}{2}\,(s_*-s)^2 ,
\label{H1II}
\eeqn
where we have used 
\bdis
\alpha-\alpha_*=\ln(x/x_*)=\tfrac32(s_*-s). 
\edis
This agrees with direct integration of (\ref{H1s}) with $H_0\simeq v$. At the end of Region~II, $s_*-s=2s_*=\tfrac23N_e$, the condensate is reduced by
\beqn
\frac{|H_1|}{v}=\frac12\,\mathcal K\,C\,N_e^2\simeq
\begin{cases}0.044, & \mathcal K=12\lambda,\\ 0.088, & \mathcal K=24\lambda,\end{cases}
\label{reduction}
\eeqn
with $N_e=\ln(1/\Omega^2)\simeq37.9$. These values are upper bounds once the running of $\lambda$ is taken into account, since $\lambda(1/\tau)$ decreases through Region~II and becomes negative above the instability scale. The one-loop correction therefore reduces the condensate by at most about ten per cent, and the Higgs field remains in its broken phase throughout the regime of validity of the effective theory.

\section{Discussion}

In this paper I have studied the Higgs field inside a Schwarzschild black hole, including a Planck-suppressed coupling between the Higgs field and the Kretschmann scalar, the leading curvature invariant in a Ricci-flat geometry. The essential point is that the radial coordinate is timelike in the interior, so that the Higgs condensate must be evolved dynamically rather than assumed to follow the minimum of its potential. Introducing a suitable interior time coordinate, I showed that the field equation reduces to Bessel's equation in the region where the curvature becomes important, with solution $H=vJ_0(x)$, where $x\propto\sqrt\xi\,t_p/\tau$ and $\tau$ is the proper time remaining before the singularity.
\\

This solution shows that the condensate cannot respond to the curvature-induced change in its potential. Although the potential acquires the symmetric form at the critical curvature $K_c=2\lambda v^2/(\xi\ell_p^2)$, the field remains frozen at its broken-phase value, and departs from it only when the curvature becomes Planckian, at a proper time of order $\sqrt\xi\,t_p$ before the singularity. This behaviour is independent of the mass of the black hole, and it replaces the static picture of curvature-induced symmetry restoration.
\\

I then included the one-loop correction from vacuum fluctuations of the Higgs field. Treated perturbatively about the classical solution, it leads to the same Bessel equation with a source term, which can be solved exactly by variation of parameters. The correction reduces the condensate by at most $\tfrac12\mathcal K C N_e^2$, about four per cent for the radial mode and nine per cent when the Goldstone modes are included, and these values are upper bounds once the running of the Higgs self-coupling is taken into account. Vacuum fluctuations therefore do not restore the electroweak symmetry, and the Higgs field remains in its broken phase throughout the interior, within the regime of validity of effective field theory.
\\

These conclusions are subject to several limitations. The one-loop analysis uses the renormalised vacuum polarisation near the singularity computed in \cite{JiangJiang}, and includes only the Higgs sector; the contributions of the electroweak gauge bosons and the top quark have not been computed, although simple estimates suggest that they do not alter the conclusion. Close to the singularity the curvature becomes Planckian, and the behaviour of the Higgs field there can only be addressed within a theory of quantum gravity; any restoration of the electroweak symmetry inside a Schwarzschild black hole must therefore occur in this regime.
\\

The Higgs vacuum thus appears remarkably robust against gravitational curvature. Two extensions would be of interest. First, the interiors of rotating and charged black holes contain an inner (Cauchy) horizon, near which infalling fluctuations are strongly blue-shifted and the classical geometry is unstable to mass inflation; the Higgs dynamics in this region may differ considerably from the Schwarzschild case. Secondly, the same reasoning applies in cosmology: in a radiation-dominated universe $K\propto t^{-4}$, and a Higgs--Kretschmann coupling of order unity would affect the Higgs field only close to the Planck epoch.

\paragraph{Acknowledgement:} I would like to thank Prof Derek Raine (emeritus professor of the University of Leicester) for some useful discussions while preparing this paper.

\paragraph{Funding:} The author received no financial support for this research.


\newpage
\bibliographystyle{elsarticle-num}  
\bibliography{references}  

\end{document}